\documentclass[twocolumn,showpacs,preprintnumbers,amsmath,amssymb]{revtex4}

\usepackage{graphicx}
\usepackage{dcolumn}
\usepackage{bm}

\begin{document}

\preprint{}

\title{Line shapes and time dynamics of the F\"orster resonances between two Rydberg atoms in a time-varying electric field}
\author{E.~A.~Yakshina$^{1, 2}$}
\author{D.~B.~Tretyakov$^{1, 2}$}
\author{I.~I.~Beterov$^{1, 2}$}
\author{V.~M.~Entin$^{1, 2}$}
\author{C.~Andreeva$^{3, 4}$}
\author{A.~Cinins$^{4}$}
\author{A.~Markovski$^{4, 5}$}
\author{Z.~Iftikhar$^{6}$}
\author{A.~Ekers$^{4, 7}$}
\author{I.~I.~Ryabtsev$^{1, 2}$}
  \email{ryabtsev@isp.nsc.ru}
\affiliation{$^1$Rzhanov Institute of Semiconductor Physics SB RAS, 630090 Novosibirsk, Russia}
\affiliation{$^2$Novosibirsk State University, 630090 Novosibirsk, Russia}
\affiliation{$^3$Institute of Electronics, Bulgarian Academy of Sciences, 1784 Sofia, Bulgaria}
\affiliation{$^4$University of Latvia, LV-1002 Riga, Latvia }
\affiliation{$^5$Technical University of Sofia, 1000 Sofia, Bulgaria}
\affiliation{$^6$Institut d'Optique, 91127 Palaiseau, France}
\affiliation{$^7$King Abdullah University of Science and Technology, Thuwal 23955-6900, Saudi Arabia}
\date{24 October 2016}

\begin{abstract}

The observation of the Stark-tuned F\"orster resonances between Rydberg atoms excited by narrowband cw laser radiation requires usage of a Stark-switching technique in order to excite the atoms first in a fixed electric field and then to induce the interactions in a varied electric field, which is scanned across the F\"orster resonance. In our experiments with a few cold Rb Rydberg atoms we have found that the transients at the edges of the electric pulses strongly affect the line shapes of the F\"orster resonances, since the population transfer at the resonances occurs on a time scale of $\sim$100 ns, which is comparable with the duration of the transients. For example, a short-term ringing at a certain frequency causes additional radio-frequency-assisted F\"orster resonances, while non-sharp edges lead to asymmetry. The intentional application of the radio-frequency field induces transitions between collective states, whose line shape depends on the interaction strengths and time. Spatial averaging over the atom positions in a single interaction volume yields a cusped line shape of the F\"orster resonance. We present a detailed experimental and theoretical analysis of the line shape and time dynamics of the Stark-tuned F\"orster resonances ${\rm Rb}(nP_{3/2} )+{\rm Rb}(nP_{3/2} )\to {\rm Rb}(nS_{1/2} )+{\rm Rb}([n+1]S_{1/2} )$ for two Rb Rydberg atoms interacting in a time-varying electric field. 

\end{abstract}

 \maketitle

\section{Introduction}

Long-range interactions between highly excited Rydberg atoms are of interest for quantum information processing with single neutral atoms [1,2]. Resonant dipole-dipole interaction between atoms in an identical \textit{nL} Rydberg state can be implemented via Stark-tuned F\"orster resonances [3,4]. A Rydberg state should be exactly midway between two other Rydberg states of the opposite parity [Fig.~1(a)] to induce a F\"orster resonance. The electric field corresponding to the F\"orster resonance has certain values defined by the polarizabilities and energy spacing of Rydberg states in a zero field. Upon appropriate experimental conditions (highly stable and homogeneous electric field, absence of stray electric fields, weak dipole-dipole interaction), Stark-tuned F\"orster resonances can be as narrow as a few millivolts per centimeter [5].

In our earlier papers [6-8] we studied the line shape of the F\"orster resonance ${\rm Rb}(37P_{3/2} )+{\rm Rb}(37P_{3/2} )\to {\rm Rb}(37S_{1/2} )+{\rm Rb}(38S_{1/2} )$ in cold Rb Rydberg atoms excited by a broadband pulsed laser radiation via a three-photon transition. The resonance was recorded by slowly scanning a weak electric field near the resonant value of 1.79~V/cm [Fig.~1(c)]. When the field was scanned by $\sim$0.2~V/cm, the frequency of the optical transition $6S_{1/2} \to 37P_{3/2} $ on the third excitation step changed by tens of megahertz due to the Stark effect. This frequency change was unimportant for broadband pulsed radiation as the radiation line width was much broader ($\sim$10~GHz). However, an electric-field pulse was applied at the moment of laser excitation to pull out the cold Rb photoions formed due to photoionization by the intense pulsed laser radiation; otherwise, the F\"orster resonance became asymmetrical due to inhomogeneous electric fields from the photoions [8].

In our recent experiments [2,9,10] we have switched to three-photon excitation by narrowband cw lasers. This prevents the appearance of photoions, since the laser intensities used are much lower than those for pulsed excitation, and the pulling electric-field pulse is not required. However, scanning the electric field by just 50~mV/cm shifts the Rydberg level out of resonance with the laser radiation, as the cw lasers have line widths below 1 MHz [9]. In order to avoid this problem, one needs to apply a Stark-switching technique [11-14] to excite the atoms first in a fixed electric field and then to induce the interactions in a lower electric field, which is scanned across the F\"orster resonance [10,13], as shown in Fig.~1(b). 

We have found that the transients at the edges of the controlling electric pulses strongly affect the line shapes of the F\"orster resonances, since the population transfer at the resonances occurs on a time scale of $\sim$100~ns, which is comparable with the duration of the transients. For example, a short-term ringing at a certain frequency causes additional radio-frequency (rf)-assisted F\"orster resonances, while non-sharp edges lead to asymmetry (see Fig.~3). The short-time scale of the Rydberg interactions between cold atoms was also observed in Refs.~[15,16].

\begin{figure}
\includegraphics[scale=0.65]{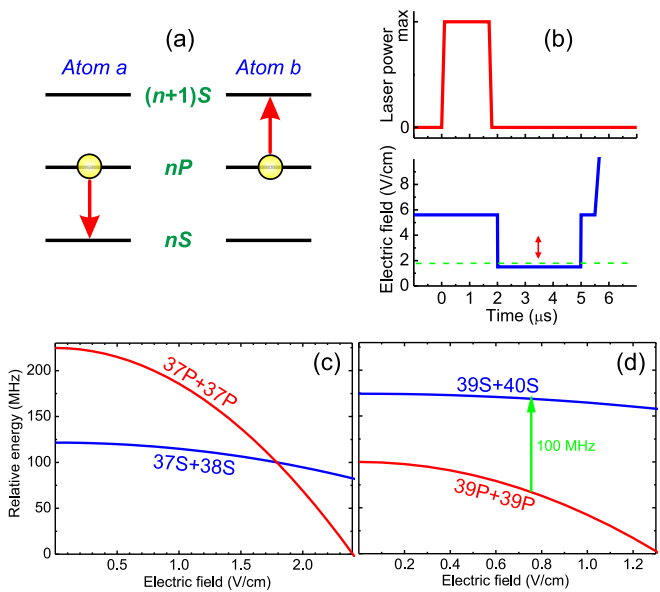}
\caption{\label{Fig1} (color online) (a) F\"orster resonance Rb(\textit{nP})+Rb(\textit{nP})$\rightarrow$Rb(\textit{nS})+Rb([\textit{n}+1]\textit{S}) for two Rb Rydberg atoms. Dipole-dipole interaction induces transitions from the initial $nP_{3/2}(|M_J|=1/2)$ state to the final $nS_{1/2}(|M_J|=1/2)$ and $(n+1)S_{1/2}(|M_J|=1/2)$ states in both atoms. (b) Timing diagram of the laser and electric-field pulses. (c) Collective states of the F\"orster resonance Rb(37\textit{P})+Rb(37\textit{P})$\rightarrow$Rb(37\textit{S})+Rb(38\textit{S}) in the dc electric field. Intersection at 1.79~V/cm corresponds to a Stark-tuned F\"orster resonance. (d)~Collective states of the F\"orster resonance Rb(39\textit{P})+Rb(39\textit{P})$\rightarrow$Rb(39\textit{S})+Rb(40\textit{S}) in the dc electric field. A dc Stark-tuned F\"orster resonance is impossible, but it can be induced by adding a $\sim$100-MHz radio-frequency field (green arrow) that binds the collective states [10].}
\end{figure}

In this paper, we present a detailed experimental and theoretical analysis of the F\"orster resonance line shapes and time dynamics in a time-varying electric field used for the Stark switching. The resonance under study is the F\"orster resonant energy transfer ${\rm Rb}(nP_{3/2} )+{\rm Rb}(nP_{3/2} )\to {\rm Rb}(nS_{1/2} )+{\rm Rb}([n+1]S_{1/2} )$ due to dipole-dipole interaction of two Rb Rydberg atoms [Fig.~1(a)] in a small single laser excitation volume of a frozen Rydberg gas. The energy detuning of this resonance, $\hbar \Delta =E(nS_{1/2} )+E([n+1]S_{1/2} )-2E(nP_{3/2} )$, is controlled by a weak dc electric field $F$. The energy shift of a Rydberg level $nL$ with nonzero quantum defect is quadratic and is defined by its polarizability, $\delta E_{nL} /\hbar =-\alpha _{nL} F^{2} /2$. The detuning is then given by

\begin{equation} \label{Eq1} 
\Delta=\Delta _{0} +(\alpha _{nP} -\frac{1}{2} \alpha _{nS} -\frac{1}{2} \alpha _{[n+1]S} )F^{2}.    
\end{equation} 
 
\noindent Here, $\Delta _{0} $ is the detuning in a zero electric field (for example, $-$103~MHz for \textit{n}=37 and +74~MHz for \textit{n}=39). The detuning can be tuned to zero for Rydberg states with $n\le 38$, as shown in Fig.~1(c) for the $37P_{3/2}(|M_J|=1/2)$ state, while for states with $n\ge 39$, the dc electric field only increases $\Delta$, as shown in Fig.~1(d) for the $39P_{3/2}(|M_J|=1/2)$ state. For the latter, a F\"orster resonance can be obtained by adding a $\sim$100~MHz rf field that induces rf transitions between collective states and compensates for the F\"orster energy defect, as  demonstrated in our recent paper [10].

An experimental and theoretical line-shape analysis of the Stark-tuned F\"orster resonances was performed in Ref.~[17] for the case of potassium Rydberg atoms colliding in a thermal atomic beam. It was found that averaging over atomic velocities leads to a cusped line shape, while resonant collisions in a velocity-selected atomic beam have nearly Lorentzian line shape. This finding agrees with what we have observed in a similar experiment on resonant Rydberg collisions in a sodium thermal atomic beam [18].

It is quite surprising that cusp-shaped F\"orster resonances are also observed in cold Rydberg atom gases, where the atoms are nearly frozen [19,20]. Line-shape analysis was performed in several papers. In Ref.~[21], it was found theoretically that the rare pair fluctuation at small spatial separation is the dominant factor contributing to the broadening of the F\"orster resonances. Our numerical Monte Carlo simulations using the Schr\"odinger's equation [7,22] and our experimental data [7,10] have shown that spatial averaging over the random positions of $N=2-5$ Rydberg atoms interacting in a single laser excitation volume results in cusp-shaped F\"orster resonances as soon as the resonances saturate and the interaction time is long enough. Cusp-shaped F\"orster resonances were also observed recently at high Rb Rydberg atom density in Ref.~[23] and an analytical model has been proposed. Another recent paper [24] reports on the broadening and overlapping of the F\"orster resonances in Rb Rydberg atoms at high density. Quasi-forbidden F\"orster resonances in Cs Rydberg atoms were observed in Ref.~[25]. Finally, in Ref.~[26], microwave pump-probe experiments also demonstrated cusped line shapes observed in the spectra of dipole-dipole broadened microwave transitions in Rb Rydberg atoms. The line-shape analysis of the resonant dipole-dipole interaction thus remains a hot topic in the study of cold Rydberg atom samples.

\section{Experimental setup}

The experiments are performed with cold $^{85}$Rb atoms in a magneto-optical trap (MOT). Typically, 10$^5-10^6$ atoms are trapped in a cloud of 0.5$-$0.6 mm diameter. Our experiments feature atom-number-resolved measurement of the signals obtained from $N=1-5$ detected Rydberg atoms with a detection efficiency of 65\% [7]. It is based on a selective field ionization (SFI) detector with channel electron multiplier (CEM) and post-selection technique. We can record F\"orster resonances for up to five of the detected Rydberg atoms and compare these with the numerical Monte Carlo simulations.

The electric field for SFI is formed by two stainless-steel plates that are 1~cm apart. These plates have holes and meshes for passing the vertical cooling laser beams and the electrons to be detected. The CEM output pulses from the \textit{nS} and [\textit{nP}+(\textit{n}+1)\textit{S}] states (the two latter states have nearly identical ionizing fields and are detected together) are detected with two independent gates and sorted according to the number \textit{N} of the totally detected Rydberg atoms. The normalized \textit{N}-atom signals $S_N$ are the fractions of atoms that have undergone a transition to the final \textit{nS} state. 

\begin{scriptsize}\begin{footnotesize}\end{footnotesize}\end{scriptsize}The excitation of Rb atoms to the \textit{nP}$_{3/2}$(\textbar M$_J$\textbar=1/2) Rydberg states is realized via three-photon transition $5S_{1/2} \to 5P_{3/2} \to 6S_{1/2} \to nP_{3/2} $  [Fig.~2(a)] by means of three cw lasers modulated to form 2~$\mu $s exciting pulses at a repetition rate of 5~kHz [9]. The first excitation step is blue detuned by $\delta_1$=+92~MHz from the $5S_{1/2} (F=3)\to 5P_{3/2} (F=4)$ transition in $^{85}$Rb atoms. The second step is on-resonance ($\delta_2$=0) with the $5P_{3/2} (F=4)\to 6S_{1/2} (F=3)$ transition. By scanning the frequency of the third-step laser across the $6S_{1/2} (F=3)\to nP_{3/2} $ transition, we observe a strong narrow peak of the coherent three-photon excitation [solid arrows in Fig.~2(a)]  detuned by $\delta_3$=$-$92~MHz [Fig.~2(b)], while the incoherent three-step excitation [dashed arrows in Fig.~2(a)] is almost absent, as discussed in our papers [2,9]. The intermediate states $5P_{3/2} $ and $6S_{1/2} $ are thus almost not populated. 

Small Rydberg excitation volume of 20$-$40 $\mu $m in size is formed using crossed laser-beam geometry [6]. The effective volume size can be controlled by changing the laser intensity on the second excitation step (higher intensity gives bigger volume). The intensity is measured by observing the Autler-Townes splitting of the $6S_{1/2} $ state for the incoherent three-step excitation [small peaks at $\pm$26 MHz in Fig.~2(b)]. The small peak at $\delta_3$=0 in Fig.~2(b) is a zero reference due to the added cooling laser radiation on the first step. The blue (thin) lines in Fig.~2(b) are theoretical calculations in a four-level model taking into account the finite laser line widths [9]. Application of a weak electric field shifts and splits the three-photon resonance, as shown in Fig.~2(c). This allows us to use the Stark-switching technique to control the interaction of Rb atoms with the laser radiations.

\begin{figure}
\includegraphics[scale=0.48]{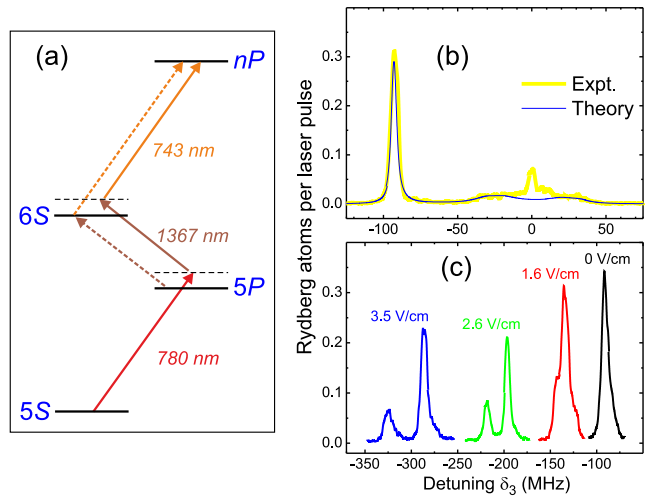}
\caption{\label{Fig2}(color online) (a)~Three-photon laser excitation of Rb(\textit{nP}) Rydberg atoms. The first step 5\textit{S}$\rightarrow$5\textit{P} is detuned by $\delta_1$=+92 MHz, while the second step 6\textit{S}$\rightarrow$\textit{nP} is tuned on-resonance $\delta_2$=0. The solid arrows indicate coherent three-photon excitation and the dashed arrows indicate incoherent three-step excitation, as discussed in Ref.~[9]. (b)~Spectrum of the three-photon laser excitation of the 37\textit{P}$_{3/2}$ state in a zero electric field. The Rabi frequencies of the intermediate transitions are $\Omega_1$=10~MHz, $\Omega_2$=26~MHz, and $\Omega_3$=2~MHz. The blue (thin) line is a theoretical calculation in a four-level model taking into account the laser line widths according to Ref.~[9]. The small peak at $\delta_3$=0 is a reference due to the added cooling laser radiation on the first step. (c)~Stark effect in the laser excitation spectrum of the 37\textit{P}$_{3/2}$ state in a weak dc electric field.}
\end{figure}

The laser intensities are adjusted to obtain about one Rydberg atom excited per laser pulse on average. A proper choice of the exciting laser polarization (along the dc electric field) provides excitation of only \textit{nP}$_{3/2}$(\textbar M$_J$\textbar=1/2) atoms from the intermediate $6S_{1/2} $ state. In this case, a single F\"orster resonance is observed, which is most appropriate for experimental and theoretical analysis.

The dc electric field is calibrated with 0.2\% uncertainty using the Stark spectroscopy of the microwave transition 37\textit{P}$_{3/2}$ $\rightarrow$ 37\textit{S}$_{1/2}$ at 80.124~GHz [6,8]. The MOT magnetic field is not switched off in order to have high repetition rate, but microwave probing on the same transition allowed us to align the excitation point to a nearly zero magnetic field [6]. The 780~nm cooling laser beams are switched off before Rydberg excitation and then switched on after the SFI detection of Rydberg atoms.

We use a Stark-switching technique [10-13] to switch the Rydberg interactions on and off, as depicted in Fig.~1(b). Laser excitation occurs during 2~$\mu $s at a fixed electric field of 5.6~V/cm. Then the field decreases to a lower value near the resonant electric field (1.79~V/cm for the 37\textit{P}$_{3/2}$ state), which acts for 3~$\mu $s or less until the field increases back to 5.6~V/cm. Then, 0.5~$\mu $s later, a ramp of the strong field-ionizing electric pulse of 200~V/cm is applied. The lower electric field is slowly scanned across the F\"orster resonance and the SFI signals are accumulated for $10^3-10^4$ laser pulses. A pulse of the rf field with variable amplitude ($0-0.5$~V/cm) and frequency ($10-100$~MHz) can be admixed to the lower dc field.

\section{Effect of the edges of the electric-field pulse}

The usage of the Stark-switching technique requires studying the effect of the finite duration of the edges and of the electric-field pulse itself on the observed line shape of the F\"orster resonance. As noted in Refs.~[15,16], population transfer at a F\"orster resonance may occur rapidly, on a time scale of $\sim$100~ns. If the edges have similar duration, these can affect the line shape, which under conditions should be symmetrical and close to Lorentzian, except for the wings of the resonance [7]. At the same time, too fast switching may cause non-adiabatic transitions between Rydberg states, so the edges must not be too short (longer than $\sim$1~ns). 

In our first experiments on the Stark switching of the F\"orster resonance ${\rm Rb}(37P_{3/2} )+{\rm Rb}(37P_{3/2} )\to {\rm Rb}(37S_{1/2} )+{\rm Rb}(38S_{1/2} )$, we revealed that when we tried to obtain the 10-ns edges, on the leading edge there appeared a 300~ns transient in the form of damped 20~MHz oscillations of the voltage. It resulted in the appearance of additional weaker and broader F\"orster resonances having a regular structure of repetitions with the frequency interval being also about 20~MHz [Fig.~3(a)]. In fact, these resonances were induced by the pulse of the radio-frequency (rf) field corresponding to the ringing.

\begin{figure}
\includegraphics[scale=0.5]{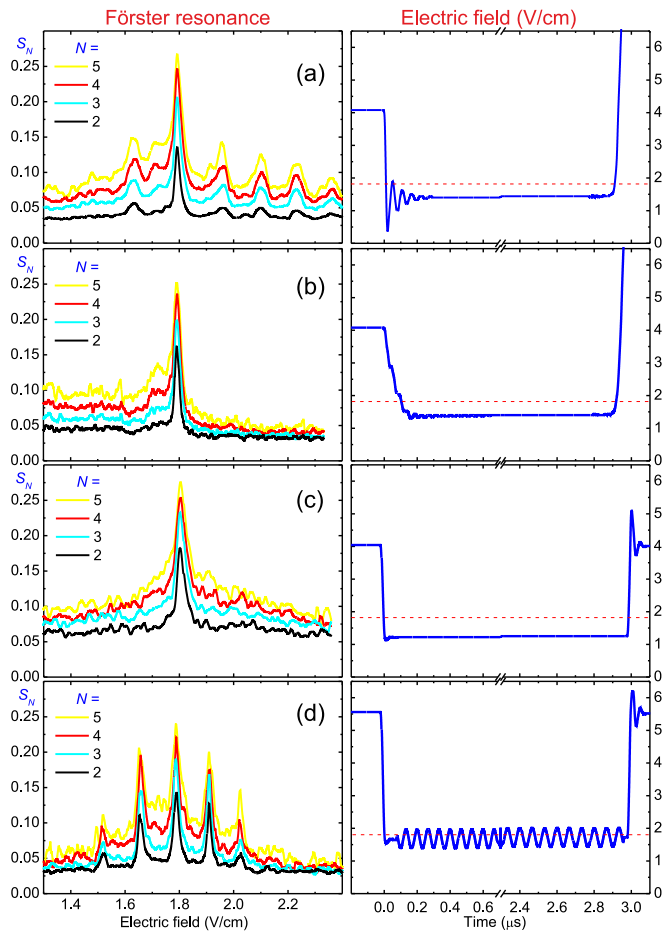}
\caption{\label{Fig3} (color online) Line shapes of the F\"orster resonance Rb(37\textit{P})+Rb(37\textit{P})$\rightarrow$Rb(37\textit{S})+Rb(38\textit{S}) recorded for $N=2-5$ detected Rydberg atoms at various time dependences of the controlling electric-field pulse shown on the right-hand panels (the dashed line indicates the resonant electric field). (a) The pulse has a short-term ringing at 20~MHz on the leading edge. (b) The ringing is smoothed using a capacitor. (c)~The pulse has a nearly square shape with short edges. (d)~A radio-frequency (rf) pulse is intentionally added to induce additional rf-assisted F\"orster resonances.}
\end{figure}

Then we suppressed the ringing and increased the leading edge duration to approximately 100~ns, while the falling edge was shorter (30~ns). The usage of such pulse with asymmetric edges has led to a noticeable asymmetry in the line shape of the F\"orster resonance [Fig.~3(b)]. This can be understood from the fact that in our experimental conditions, the population transfer at F\"orster resonance occurs mainly during the long leading edge. 

After a substantial improvement of the electric circuitry, we managed to obtain an electric pulse with short edges ($\sim$10~ns) and without any noticeable ringing or transients. This allowed us to record the symmetrical spectrum of the F\"orster resonance at Stark switching [Fig.~3(c)], which is suitable for comparison with theory.

Finally, the intentional application of a 100~mV rf-field at 15~MHz induced additional F\"orster resonances [Fig.~3(d)], which are  rf-assisted resonances of various orders, as shown in the scheme in Fig.~6(a) for the energy levels of the initial 37\textit{P}+37\textit{P} and final 37\textit{S}+38\textit{S} collective states in dc electric field. We studied these resonances experimentally in Ref.~[10].

In all cases considered above, the F\"orster resonance amplitudes and widths grow as the number of atoms \textit{N} increases from 2 to 5, due to an increase in the total interaction energy and density of Rydberg atoms in the same laser excitation volume. The cusp-shaped F\"orster resonances are formed as soon as the resonances saturate and the interaction time is long enough. This observation agrees with our previous theoretical Monte Carlo simulations of the F\"orster resonances for $N=2-5$ Rb Rydberg atoms randomly placed in a single excitation volume [7,22].

\section{Time dynamics of the F\"orster resonances}

\subsection{Theory with density-matrix equations}

The next issue to be analyzed is the dependence of the line shape and amplitude of the F\"orster resonance on the interaction time \textit{t} which is set by the length of the controlling electric-field pulse [Fig.~3(c)]. We consider an example of the Stark-tuned F\"orster resonance ${\rm Rb}(37P_{3/2} )+{\rm Rb}(37P_{3/2} )\to {\rm Rb}(37S_{1/2} )+{\rm Rb}(38S_{1/2} )$ for two Rb Rydberg atoms [see Fig.~1(a) for the energy level and transition scheme] randomly placed in a single excitation volume. The initial energy detuning $\hbar \Delta =[E(37S_{1/2} )+E(38S_{1/2} )-2E(37P_{3/2} )]$ in a zero electric field is $-$103 MHz. For the laser-excited 37\textit{P}$_{3/2}$(\textbar M$_J$\textbar =1/2) Rydberg atoms a single F\"orster resonance is observed at 1.79~V/cm, as shown in Fig.~3(c). 

Our F\"orster resonance induces transitions between Rydberg states with $\Delta {\rm M}_{J} =0$. This corresponds to the \textit{z}-oriented dipoles, and the operator of the dipole-dipole interaction is

\begin{equation} \label{Eq2} 
\hat{V}=\frac{\hat{d}_{a} \hat{d}_{b} }{4\pi \varepsilon _{0} } \left[\frac{1}{R_{ab}^{3} } -\frac{3\, \, Z_{ab}^{2} }{R_{ab}^{5} } \right],   
\end{equation} 

\noindent where $\hat{d}_{a,b} $ are the \textit{z} components of the dipole-moment operators of the two interacting atoms \textit{a} and \textit{b}, $Z_{ab} $ is the \textit{z }component of the vector connecting the two atoms $\mathbf{R}_{ab} $ (\textit{z} axis is chosen along the dc electric field), and $\varepsilon_0$ is the dielectric constant.

In order to calculate the time evolution of the populations in the two Rydberg atoms interacting via a F\"orster resonance, the simplest way is to solve the Schr\"odinger's equation for two interacting atoms. For two motionless atoms, this equation is solved analytically. Then the F\"orster resonance line shape and time dynamics can be obtained by Monte Carlo spatial averaging of the random atom positions in the laser excitation volume, as we did in Refs.~[7,22]. 

However, when comparing the experimental and theoretical line shapes in Ref.~[7], we have found that in the experiment, we observed additional broadening of the F\"orster resonance, which was not described by the Schr\"odinger's equation for three-level Rydberg atoms shown in Fig.~4(a). The additional broadening was due to the unresolved hyperfine structure of Rydberg states not taken into account in theory ($\sim$500~kHz) and due to instability of the controlling dc electric field ($\sim$5~mV/cm). Accounting for the hyperfine structure in the theory would strongly increase the number of equations for all collective states related to the F\"orster resonance, so that any analytical solution would be impossible, while numerical simulations would require a long time.

Therefore, the question about a simple theory correctly describing the line shapes of our F\"orster resonances observed for arbitrary interaction time remains open. In this paper, instead of the Schr\"odinger's equation, we elaborate a more adequate density-matrix model, which takes into account the additional broadening phenomenologically. Our considerations are limited to the case of \textit{N}=2 interacting Rydberg atoms, when approximate analytical solutions can be found and compared with our experimental data.

Let us denote the lower energy state 37\textit{S} as state 1, the middle state 37\textit{P}$_{3/2}$ as state 2, and the upper state 38\textit{S} as state 3, as shown in Fig.~4(a). Then, for two interacting Rydberg atoms with dipole-dipole matrix element \textit{V} given by Eq.~\eqref{Eq2}, ${\left| 22 \right\rangle} $ is the initial collective state populated by a short laser pulse at $t=0$, and ${\left| 13 \right\rangle} $ and ${\left| 31 \right\rangle} $ are the two equally populated final states having a small energy detuning $\Delta $ from state ${\left| 22 \right\rangle} $ [Fig.~4(b)]. We ignore the other collective Rydberg states ${\left| 21 \right\rangle} ,{\left| 12 \right\rangle} ,{\left| 23 \right\rangle} ,{\left| 32 \right\rangle} $, since they have large energy detunings from state ${\left| 22 \right\rangle} $ and are not populated at F\"orster resonance. 

\begin{figure}
\includegraphics[scale=0.62]{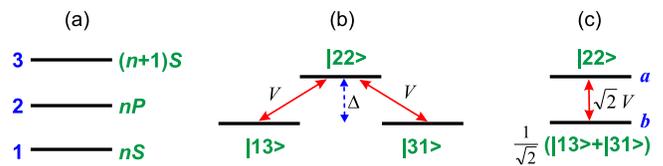}
\caption{\label{Fig4} (color online) (a) Rydberg states related to the F\"orster resonance Rb(\textit{nP})+Rb(\textit{nP})$\rightarrow$Rb(\textit{nS})+Rb((\textit{n}+1)\textit{S}). (b) Collective states of two Rydberg atoms interacting via F\"orster resonance with dipole-dipole matrix element \textit{V}. The energy defect of the F\"orster resonance $\Delta$ is controlled by the electric field. (c) Effective two-level system can replace the three-level system in the theoretical calculations, as collective states \textbar 13$>$ and \textbar 31$>$ are equivalent and behave identically. The reduced dipole-dipole matrix element is now $\sqrt{2} \; V$.}
\end{figure}

A simpler effective two-level system, shown in Fig.~4(c), with a reduced dipole-dipole matrix element $\sqrt{2} \; V$ can replace the three-level system of Fig.~4(b) in the theoretical calculations, because collective states ${\left| 13 \right\rangle} $ and ${\left| 31 \right\rangle} $ are equivalent and behave identically. We now denote state ${\left| 22 \right\rangle} $ as state $a$ in the effective two-level system and the symmetric composite state $({\left| 13 \right\rangle} $+${\left| 31 \right\rangle)/\sqrt{2}} $ as state $b$ in Fig.~4(c). The other, antisymmetric composite state  $({\left| 13 \right\rangle} $-${\left| 31 \right\rangle)/\sqrt{2}} $ is not considered, as it is unaffected by the operator of the dipole-dipole interaction $\hat{V}$ and can be excluded from the analysis.

The most convenient form of equations for the density-matrix elements $\rho_{ij}$ is the optical Bloch equations [27], where the complex exponents are excluded by a replacement of variables, and the detuning $\Delta$ is present only in the expressions for the reduced coherences ($i\neq j$) and is absent in the expressions for the populations ($i=j$):

\begin{equation} \label{Eq3} 
\begin{array}{l} {\dot{\rho }_{aa} =i\sqrt{2} \; \Omega \; \left(\rho _{ab} -\rho _{ba} \right),} \\ {\dot{\rho }_{bb} =i\sqrt{2} \; \Omega \; \left(\rho _{ba} -\rho _{ab} \right),} \\ {\dot{\rho }_{ab} =-(i\Delta +\Gamma /2)\rho _{ab} +i\sqrt{2} \; \Omega \; \left(\rho _{aa} -\rho _{bb} \right),} \\ {\dot{\rho }_{ba} =(i\Delta -\Gamma /2)\rho _{ba} +i\sqrt{2} \; \Omega \; \left(\rho _{bb} -\rho _{aa} \right).} \end{array} 
\end{equation} 

\noindent Here, $\Omega =V/\hbar $ is the matrix element of the dipole-dipole interaction in circular frequency units. In Eqs.~\eqref{Eq3}, we neglect the spontaneous and blackbody-radiation-induced depletion of the Rydberg states, as their effective lifetimes are tens of microseconds [28] and the related contribution to the F\"orster resonance width is just a few kilohertz, while in the experiments we typically observe F\"orster resonances of more than 1~MHz width. 

In order to take into account the additional broadening $\Gamma$ phenomenologically, Eqs.~\eqref{Eq3} are modified using the method we have applied previously in Ref.~[9] to account for the finite laser linewidths in a four-level theoretical model of the three-photon laser excitation of Rydberg states. In the equations for the coherences, we add the terms with $\Gamma $/2 in order to introduce additional coherence decay. This method is called the phase-diffusion model and it describes the case where the driving radiation (or interaction) has random phase fluctuations but no amplitude fluctuations [29]. The validity of this model was grounded in Ref.~[30]. Note that the noise spectrum in this model has a Lorentzian shape, which is not necessarily the case in the experiments. 

The two-atom signal $S_2$ measured in our experiments is a fraction of Rydberg atoms in the final state 37\textit{S} or a population of the final state 37\textit{S} per atom, which is calculated for the interaction time \textit{t} as

\begin{equation} \label{Eq4} 
S_{2} =\frac{1}{2} \rho _{bb} (t). 
\end{equation} 

\noindent In order to find $\rho _{bb} (t)$, Eqs.~\eqref{Eq3} can be solved analytically with the initial conditions $\rho _{aa}(0)=1;\; \rho _{bb}(0)=\rho _{ab}(0)=\rho _{ba}(0)=0$. Their solution reduces to finding the roots of a cubic equation, if $\Omega$, $\Delta$, and $\Gamma$ are independent of time (see Appendix A). The resulting general analytical expressions are rather complicated, however, and in what follows we consider the analytical solutions only for some particular cases.

\subsection{Theory for the F\"orster resonance amplitude}

The exact analytical solution of Eqs.~\eqref{Eq3} for the time evolution of the F\"orster resonance amplitude $S_{2} (\Delta =0)$ is given by

\begin{equation} \label{Eq5} 
\begin{array}{c} {\displaystyle S_{2} (\Delta =0)=\frac{1}{4} -\frac{1}{4} {\rm e}^{-\Gamma t/4} \left[{\rm \; ch}\left(\sqrt{\Gamma ^{2} /16-8\Omega ^{2} } \; t\right)\right. } \\ \\ {\left. +\displaystyle \frac{\Gamma /4}{\sqrt{\Gamma ^{2} /16-8\Omega ^{2} } } {\rm \; sh}\left(\sqrt{\Gamma ^{2} /16-8\Omega ^{2} } \; t\right)\right].} \end{array} 
\end{equation} 

\noindent At the weak dipole-dipole interaction $8\Omega ^{2} \ll\Gamma ^{2} /16$, Eq.~\eqref{Eq4} reduces to

\begin{equation} \label{Eq6} 
S_{2}^{weak} (\Delta =0)\approx \frac{1}{4} \left[1-{\rm e}^{-16\Omega ^{2} t/\Gamma } \right]. 
\end{equation} 

\noindent The amplitude slowly goes to its steady-state value 1/4 as \textit{t} increases. For the strong dipole-dipole interaction $8\Omega ^{2} \gg\Gamma ^{2} /16$, the damped Rabi-like oscillations appear, while the steady-state value is also 1/4:

\begin{equation} \label{Eq7} 
S_{2}^{strong} (\Delta =0)\approx \frac{1}{4} \left[1-{\rm e}^{-\Gamma t/4} \cos \left(2\sqrt{2} \; \Omega t\right)\right]. 
\end{equation} 

\noindent This is what can be observed with two Rydberg atoms, interacting in two spatially separated optical dipole traps, as in the experiments in Refs.~[31,32]. At $\Gamma =0$, the oscillations would never end, since there is no decoherence for the two motionless atoms considered here. But, in fact, $\Gamma$ cannot be zero as it is ultimately limited to a few kilohertz by the finite lifetimes of the Rydberg states [28].

In the realistic experiments, the atom positions are not fixed. The interaction term $\Omega$ in Eqs.~\eqref{Eq6} and \eqref{Eq7} has a fluctuating value that results in decoherence and washing out the Rabi-like oscillations in Eq.\eqref{Eq7}, even at $\Gamma =0$. To calculate the resonance amplitude measured in our experiments, Eqs.~\eqref{Eq6} and \eqref{Eq7} should be averaged over all possible spatial positions of the two interacting atoms in the excitation volume, which is formed by the two intersecting laser beams. In Ref.~[22] we have shown that the averaging can be done using the nearest-neighbor probability distribution [33,34] with the average distance between nearest-neighbor atoms $R_{0} \approx \left[3/(4\pi n_{0} )\right]^{1/3} $ at volume density \textit{n$_0$}. Using this method, we find the approximate analytical solutions to the averaged amplitudes of Eqs.~\eqref{Eq6} and \eqref{Eq7} (see Appendix B):

\begin{equation} \label{Eq8} 
<S_{2}^{weak} (\Delta =0)>\approx \frac{1}{4} \left(1-{\rm e}^{-\left[0.44\Omega _{0}^{2} \; t/\Gamma \right]^{1/3} } \right),  
\end{equation} 

\begin{equation} \label{Eq9} 
<S_{2}^{strong} (\Delta =0)>\approx \frac{1}{4} \left(1-{\rm e}^{-0.55\Omega _{0} \; t-\Gamma t/4} \right),  
\end{equation} 

\noindent where 

\noindent 

\[\Omega _{0} =\frac{\sqrt{2} d_{1} d_{2} }{4\pi \varepsilon _{0} \hbar R_{0}^{3} } \] 

\noindent is the orientation-averaged reduced interaction matrix element at the average distance $R_{0} $. Here, $d_{1} $ and $d_{2} $ are the dipole moments of transitions ${nP_{3/2} \left({\rm M}_{J} =1/2\right)} \to {nS_{1/2} \left({\rm M}_{J} =1/2\right) } $ and ${nP_{3/2} \left({\rm M}_{J} =1/2\right)} \to {(n+1)S_{1/2} \left({\rm M}_{J} =1/2\right) } $. The fitting coefficients 0.44 and 0.55 have been found from the numerical simulations in the same way as in Ref.~[22] for a cubic interaction volume (see Appendix B). 

The time dependences of the averaged amplitudes in Eqs.~\eqref{Eq8} and \eqref{Eq9} significantly differ from those in Eqs.~\eqref{Eq6} and \eqref{Eq7} for two spatially fixed atoms. This should be taken into account when theory is compared with experiment.

\subsection{Theory for the F\"orster resonance line shape}

The analytical expressions for the line shape of the two-atom F\"orster resonances at $\Delta \ne 0$ are much more complicated than Eqs.~\eqref{Eq5}-\eqref{Eq7}. In the case when $\rho _{bb} \ll\rho _{aa} $ (weak interaction or short interaction time), the line shape for two frozen Rydberg atoms is approximately given by

\begin{equation} \label{Eq10} 
\begin{array}{c} {S_{2}^{weak} \approx \displaystyle\frac{2\Omega ^{2} }{\Delta ^{2} +\Gamma ^{2} /4} \left(\frac{\Gamma t}{2} +\frac{\Delta ^{2} -\Gamma ^{2} /4}{\Delta ^{2} +\Gamma ^{2} /4} \times \right. } \\ \\\displaystyle{\left. \left[1-{\rm e}^{-\Gamma t/2} \cos \left(\Delta t\right)\right]-\frac{\Delta \Gamma }{\Delta ^{2} +\Gamma ^{2} /4} {\rm e}^{-\Gamma t/2} \sin \left(\Delta t\right)\right).} \end{array} 
\end{equation} 

\noindent Our numerical simulations have shown that this formula is quite precise (error less than 10\%) for arbitrary $\Omega$, $\Gamma$, $\Delta$, and \textit{t} as far as $S_{2} <0.1$. In fact, this is a general formula for the line shape of weak F\"orster resonances in two frozen Rydberg atoms. For the short interaction time $\Gamma t/2\ll1$, the line shape is given by just a Fourier transform of the short interaction pulse:

\begin{equation} \label{Eq11} 
S_{2}^{weak} \approx \frac{4\Omega ^{2} }{\Delta ^{2} } \sin ^{2} \left(\Delta t/2\right). 
\end{equation} 

\noindent Its full width at half maximum (\textit{FWHM}) in circular frequency units is $FWHM\approx 2\pi /t$. For the long interaction time $\Gamma t/2\gg 1$, the line shape is given by a Lorentz profile with $FWHM\approx \Gamma $,

\begin{equation} \label{Eq12} 
S_{2}^{weak} \approx \frac{\Omega ^{2} }{\Delta ^{2} +\Gamma ^{2} /4} \Gamma t \;. 
\end{equation} 

In the opposite case of strong interaction $8\Omega ^{2} \gg\Gamma ^{2} /16$, the line shape is approximately given by

\begin{equation} \label{Eq13} 
\begin{array}{c} {S_{2}^{strong} \approx \displaystyle\frac{1}{4} -\frac{\Delta ^{2} /4}{8\Omega ^{2} +\Delta ^{2} } {\rm e}^{-\frac{4\Omega ^{2} }{8\Omega ^{2} +\Delta ^{2} } \Gamma t} -} \\ \\ {\displaystyle\frac{2\Omega ^{2} }{8\Omega ^{2} +\Delta ^{2} } {\rm e}^{-\frac{4\Omega ^{2} +\Delta ^{2} }{8\Omega ^{2} +\Delta ^{2} } \Gamma t/2} \cos \left(\sqrt{8\Omega ^{2} +\Delta ^{2} } \; t\right).} \end{array} 
\end{equation} 

\noindent This formula describes the saturation of the F\"orster resonance accompanied by the damped Rabi-like oscillations. As \textit{t} or $\Omega$ grow, there appears a flat-top contour with the width of the flat-top part $4\sqrt{2} \Omega $, while the resonance wings are close to a Lorentzian. At $t\to \infty $, Eq.~\eqref{Eq13} gives $S_{2}^{strong} \to 1/4$ independently of $\Delta$. This is because the steady-state solutions $\rho _{aa} =\rho _{bb} =1/2;\; \rho _{ab} =\rho _{ba} =0$ of Eqs.~\eqref{Eq3} are independent of $\Delta$ if $\Gamma \ne 0$. For large $\Delta$, however, the time evolution is very slow, as described by the second term in Eq.~\eqref{Eq13}.

If $\Gamma =0$, the line shape is given by the undamped Rabi-like oscillations for arbitrary \textit{t},

\begin{equation} \label{Eq14} 
S_{2}^{strong} \approx \frac{4\Omega ^{2} }{8\Omega ^{2} +\Delta ^{2} } \sin ^{2} \left(\frac{1}{2} \sqrt{8\Omega ^{2} +\Delta ^{2} } \; t\right). 
\end{equation} 

\noindent The resonance width is either $FWHM\approx 2\pi /t$ at $t\ll\Omega ^{-1} $ or $FWHM\approx 4\sqrt{2} \Omega $ at $t\gg \Omega ^{-1} $.

The position-averaged line shape of the F\"orster resonance for two atoms in a single interaction volume can be obtained analytically only for Eqs.~\eqref{Eq13} or \eqref{Eq14}, because they correctly describe the saturation of resonances, otherwise the integral in the averaging diverges at short interaction distances. In Ref.~[23], the averaging has been performed for the Lorentz pre-factor of Eq.~\eqref{Eq14} and an analytical formula has been obtained in terms of the standard sine and cosine integrals. This formula predicts a cusp-shaped F\"orster resonance at high Rydberg atom density, but it does not provide the time dependence of the amplitude and line shape.

In our averaging here, we use not the Lorentz pre-factor but Eq.~\eqref{Eq13} with $\Gamma \ne 0$, as discussed in Appendix C. The averaged line shape turns out to also be a cusp, which is approximately described by the following formula:

\begin{equation} \label{Eq15} 
\left\langle S_{2}^{strong} \right\rangle \approx \frac{1}{4} \left[1-\exp \left(-\left\{\frac{0.44\; \Omega _{0}^{2} \Gamma t}{a^{2} \Delta ^{2} +\Gamma ^{2} } \right\}^{1/3} \right)\right]. 
\end{equation} 

\noindent The fitting coefficients 0.44 and \textit{a} have been obtained by comparing Eq.~\eqref{Eq15} with the numerical simulations of Eqs.~\eqref{Eq3} averaged over a cubic interaction volume for the F\"orster resonance Rb(37\textit{P})+Rb(37\textit{P})$\rightarrow$Rb(37\textit{S})+Rb(38\textit{S}). At $\Gamma /(2\pi )\sim 1$~MHz, we have found that $a\approx 2$ for $\Omega _{0} /(2\pi )>1$~MHz and $t>10\; \mu {\rm s}$; $a\approx 2\sqrt{2} $ for $\Omega _{0} /(2\pi )\sim 0.1-0.5$~MHz and $t\sim 1-5\; \mu {\rm s}$; and $a\approx 4$ for $\Omega _{0} /(2\pi )<0.1$~MHz and $t\sim 1-5\; \mu {\rm s}$. In all intermediate cases, \textit{a} should be varied to fit experimental or numerical data. 

Equation \eqref{Eq15} is thus a general formula that describes the cusped line shape of the F\"orster resonance for two Rydberg atoms randomly positioned in a single interaction volume. It is valid for a broad range of all parameters, and in addition it describes the time dynamics of the F\"orster resonance for the interaction times $t>1\; \mu {\rm s}$. In particular, it gives the same result as Eq.~\eqref{Eq8} for the averaged resonance amplitude, although Eq.~\eqref{Eq8} was derived for the weak dipole-dipole interaction, while Eq.~\eqref{Eq15} was derived from Eq.~\eqref{Eq13} describing the strong interaction. This can be explained by the fact that upon spatial averaging, the interactions between distant atoms are weak in most cases, as discussed in several other papers [17,21,23].

The formation of the cusp-shaped resonance upon spatial averaging can be understood from the fact that at zero detuning, the interaction is long range (resonant dipole-dipole) and it is effective even for the distant atoms, while for nonzero detuning the interaction is short range (van der Waals) and it is much weaker for the distant atoms [10]. Therefore, upon spatial averaging, the resonance wings are reduced stronger than the resonance center, and this finally results in the cusped line shape.

Equation \eqref{Eq15} allows us also to calculate the \textit{FWHM} of the cusped F\"orster resonance: 

\begin{equation} \label{Eq16} 
FWHM\approx \frac{2}{a} \sqrt{\frac{0.44\; \Omega _{0}^{2} \Gamma t}{\left[\ln \frac{2}{1+\exp \left(-\left\{\frac{0.44\; \Omega _{0}^{2} t}{\Gamma } \right\}^{1/3} \right)} \right]^{3} } -\Gamma ^{2} }\; .   
\end{equation} 

\noindent For the weak resonance $(0.44\Omega _{0}^{2} t/\Gamma )^{1/3} \ll1$, the width is $FWHM\approx 5.3\Gamma /a$. For the strong resonance $(0.44\Omega _{0}^{2} t/\Gamma )^{1/3} \gg1$ the width is $FWHM\approx \Omega _{0} \sqrt{5.3\Gamma t} /a$. We thus see that at the strong interaction, the line width depends on time and grows with \textit{t}. This finding is quite unusual, because it is generally believed that at the strong interaction the line width is defined only by $\Omega _{0} $ [17,21,23]. This is a specific feature of the spatial averaging in a disordered atom ensemble. It stems from the fact that upon spatial averaging the third term in Eq.~\eqref{Eq13} averages to nearly zero, while the second term remains significant at $\Gamma \ne 0$ and fully defines the averaged line shape in Eq.~\eqref{Eq15}.

The only drawback of Eq.~\eqref{Eq15} is that it is invalid for the short interaction times ($t<1\; \mu {\rm s}$), when the Fourier broadening dominates. The Fourier broadening is described by the third term in Eq.~\eqref{Eq13}, which cannot be spatially averaged analytically with its time-dependent part, so numerical calculations are required. For the analytical estimates, the averaged Fourier broadening is described with a certain accuracy by Eqs.~\eqref{Eq10} or \eqref{Eq11} if $2\Omega ^{2} $ is replaced by $\Omega _{0}^{2} $.

We note that the above theoretical considerations are valid only for two interacting Rydberg atoms, when analytical solution of the density-matrix equation is possible. For the larger \textit{N} numerical simulations should be used [22].

\subsection{Experimental F\"orster resonance line shapes}

Figure 5 presents the line shapes of the F\"orster resonance Rb(37\textit{P})+Rb(37\textit{P})$\rightarrow$Rb(37\textit{S})+Rb(38\textit{S}) recorded in our experiment for \textit{N}=2 detected Rydberg atoms at various \textit{t} . The interaction time \textit{t} is set by the length of the square-shaped controlling electric-field pulse shown in Fig.~3(c).

\begin{figure}
\includegraphics[scale=0.58]{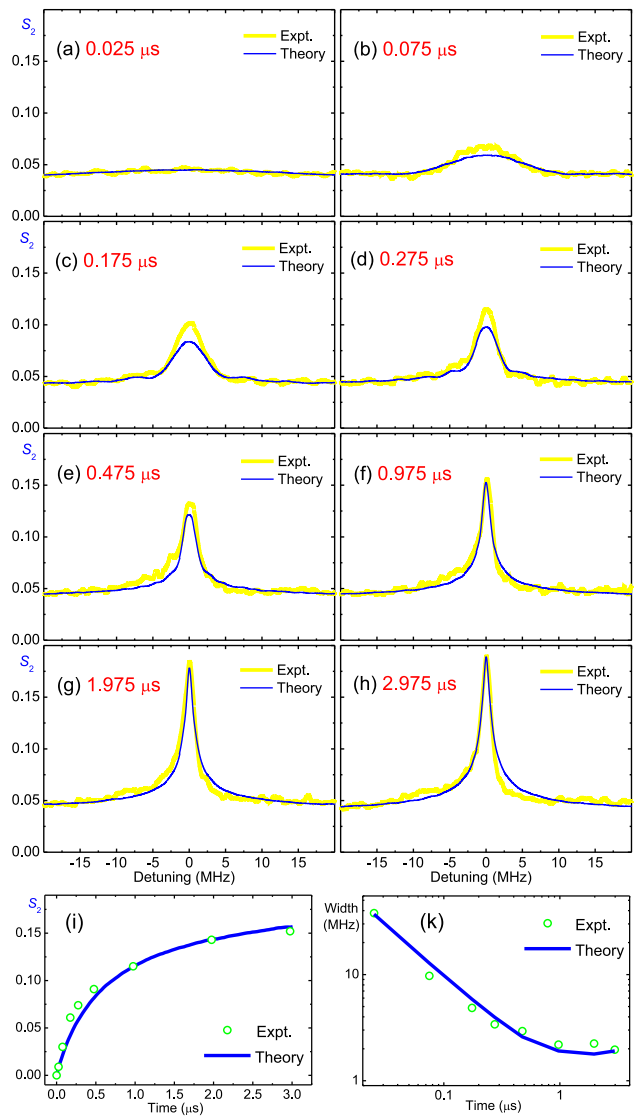}
\caption{\label{Fig5} (color online) (a)-(h) Comparison between experimental and theoretical line shapes of the F\"orster resonance Rb(37\textit{P})+Rb(37\textit{P})$\rightarrow$Rb(37\textit{S})+Rb(38\textit{S}) for two Rydberg atoms at various interaction times. (i)~Comparison between experiment and theory for the resonance amplitude at various interaction times. (k)~Comparison between experiment and theory for the resonance width at various interaction times.}
\end{figure}

At the long interaction times (1.975 and 2.975~$\mu$s), the resonance shape and width in Figs. 5(g) and 5(h) almost do not change with \textit{t} . This means that the resonance takes its stationary form, and only its amplitude slowly grows with \textit{t} according to Eq.~\eqref{Eq8}. The resonance is cusp shaped, in agreement with Eq.~\eqref{Eq15}. The observed small asymmetry in the red wing of the resonance is presumably due to some imperfection of the electric-field pulse in Fig.~3(c) because the electric field slightly varies during the pulse and its edges are not absolutely perfect. 

The narrowest resonance appears at \textit{t}=2.975~$\mu$s. Its width in the electric-field scale is 16~mV/cm, corresponding to 1.9~MHz in the detuning scale. The detuning scale is obtained using the calculated polarizabilities of the collective Rydberg states in Fig.~1(c). Our attempts to obtain a narrower resonance by increasing \textit{t} were not successful.  Since the Fourier width of the interaction pulse is $t^{-1} \approx \; 0.34$~MHz and the estimated average interaction energy is $\Omega _{0} /(2\pi )\sim 0.25$~MHz, they cannot be fully responsible for the 1.9~MHz resonance width. Therefore, there is an additional broadening $\Gamma /(2\pi )\sim 1$~MHz, apparently due to the unresolved hyperfine structure and parasitic ac electric fields. This observation agrees with what we have observed in our previous experiments [7,8]. 

The Fourier broadening of the resonances is demonstrated in Figs. 5(a)-5(e) at the shorter interaction times ($<1$~$\mu$s). The Fourier width of the interaction pulse is significant at short \textit{t }and the F\"orster resonance broadens according to Eq.~\eqref{Eq11}, while its amplitude decreases. 

In Figs.~5(a)-5(h), we also compare the experimental two-atom spectra at various \textit{t} with the line shapes numerically calculated with Eqs.~\eqref{Eq3} and averaged over the random positions of two atoms in a cubic interaction volume. The fitting parameters in the theory were the volume size 26$\times$26$\times$26~$\mu$m$^3$ and the additional broadening $\Gamma /(2\pi )=0.5$~MHz. We also added a background signal $S_{2} \approx 0.037$ that appears in the experiments due to parasitic transitions between Rydberg states induced by a 300~K blackbody radiation (BBR) [28]. These parameters allowed us to fit the time dependences of the experimental spectra, both for the amplitude [Fig.~5(i)] and width [Fig.~5(k)] of the two-atom F\"orster resonance. The amplitude and width are measured with respect to the BBR-induced background signal level.

The time dependence of the amplitude in Fig.~5(i) for $t>1$~$\mu$s is also well fit by Eq.~\eqref{Eq8} at $\Omega _{0} /(2\pi )=0.25$~MHz and $\Gamma /(2\pi )=0.5$~MHz. The time dependence of the width in Fig.~5(k) for $t<1$~$\mu$s is mainly represented by the Fourier transform width $FWHM/(2\pi )\approx 1/t$. For $t>1 \mu$s, the width becomes nearly constant and can be estimated with Eq.~\eqref{Eq16}. For $t=2.975 \mu$s, the width calculated with Eq.~\eqref{Eq16} is 1.46~MHz. If we add the Fourier width of 0.34, the total width of 1.8~MHz is close to the experimental width of 1.9~MHz. 

One can conclude that our simple theoretical density-matrix model works well and provides correct analytical and numerical results for the time dynamics and cusped line shapes of the F\"orster resonances in two randomly positioned Rydberg atoms interacting in a single laser-excitation volume. Good agreement between experiment and theory confirms that any unresolved (hyperfine or Zeeman) structure or parasitic ac Stark broadening of the F\"orster resonances can be accounted for theoretically by a single parameter $\Gamma$. This significantly reduces the number of equations and simplifies the calculations of the line shapes and time dynamics of the F\"orster resonances.

\section{Radio-frequency-assisted F\"orster resonances}

As the rf-induced F\"orster resonances can be used to enhance the interactions between Rydberg atoms at long distances and to tune the van der Waals to resonant dipole-dipole interaction, we have performed the experiments where rf field of various frequencies and amplitudes was intentionally added. Our first experimental results were published in Ref.~[10]. Here we analyze in more detail the main features of the rf-assisted F\"orster resonances and provide their comparison with theory.

\subsection{Theory of rf-assisted F\"orster resonances}

The physical interpretation of the rf-assisted F\"orster resonances was given in several papers [4,5,35,36]. A comprehensive theoretical analysis can be found in Ref.~[36]. Several features should be emphasized. 

First, the rf field induces single- and multi-photon transitions between collective states, as shown by the arrows in Fig.~6(a) for the "accessible" Stark-tuned F\"orster resonance Rb(37\textit{P})+Rb(37\textit{P})$\rightarrow$Rb(37\textit{S})+Rb(38\textit{S}) and in Fig.~6(b) for the "inaccessible" F\"orster resonance Rb(39\textit{P})+Rb(39\textit{P})$\rightarrow$Rb(39\textit{S})+Rb(40\textit{S}), which cannot be Stark-tuned by the dc electric field [10]. The rf photons of frequency $\omega $ compensate for the F\"orster energy defect $\Delta$ when it has values that are multiples of $\omega $, i.e., $\Delta =m\omega $, with \textit{m} being an integer. Multi-photon rf transitions can be driven if the rf field is strong enough. Similar F\"orster resonances between collective Rydberg states can also be induced by microwave fields [37].

Second, rf-assisted F\"orster resonances can also be explained in terms of the Floquet sidebands induced by a periodic perturbation of the Rydberg energy levels by the rf electric field due to the Stark effect [5,35,36]. Such sidebands were observed experimentally in the laser-excitation spectra of cold Rb Rydberg atoms and reported in our paper [10], as well as in several other experiments with Rydberg atoms in the vapor cells [38,39] and atomic beams [40,41]. Following Ref.~[36], one should consider the Stark effect in a composite electric field consisting of dc and rf parts, 

\begin{equation} \label{Eq17} 
F=F_{dc} +F_{rf} \; \cos (\omega t).    
\end{equation} 

\noindent Then the time-dependent energy shift of a Rydberg level with nonzero quantum defect is 

\begin{equation} \label{Eq18} 
\begin{array}{l} {\delta E_{nL} /\hbar =-\frac{1}{2} \alpha _{nL} [F_{dc}^{2} +\frac{1}{2} F_{rf}^{2} +} \\ \\{2F_{dc} F_{rf} \cos (\omega t)+\frac{1}{2} F_{rf}^{2} \cos (2\omega t)]}\;. \end{array} 
\end{equation} 

\noindent The term $F_{rf}^{2} /2$ in the brackets is responsible for the ac Stark shift of the Rydberg level in the rf field [36]. The terms with $\cos (\omega t)$ and $\cos (2\omega t)$ drive the transitions between collective states if $\Delta =m\omega $. 

The Floquet approach gives the eigenenergies of a Rydberg atom in a composite dc+rf field [36] as an infinite number of energy sidebands separated by $\omega $ [36]. For the Rydberg states experiencing quadratic Stark effect the relative amplitudes $a_{nL,m} $ of the wave functions of the Floquet sidebands are described by the generalized Bessel functions,

\begin{equation} \label{Eq19} 
a_{nL,m} =\sum _{k=-\infty }^{\infty }J_{m-2k} \left(\frac{\alpha _{nL} F_{dc} F_{rf} }{\omega } \right) \; J_{k} \left(\frac{\alpha _{nL} F_{rf}^{2} }{8\; \omega } \right),   
\end{equation} 

\noindent while the wave function of the \textit{nL} Rydberg state is

\begin{equation} \label{Eq20} 
\Psi _{nL} (r,t)=\psi _{nL} (r)e^{i\alpha _{nL} (F_{dc}^{2} +F_{rf}^{2} /2)t/2} \sum _{m=-\infty }^{\infty }a_{nL,m} {\rm e}^{im\omega t}  .  
\end{equation} 

The rf-assisted F\"orster resonances arise for the Floquet sidebands that satisfy the resonance condition $\Delta =m\omega $ and intersect at some particular values of the dc electric field. Such resonances can also be observed when the rf-frequency is scanned and rf transitions of various orders are consequently induced [5]. At $F_{dc} =0$, the odd sidebands disappear according to Eq.~\eqref{Eq18} since only $J_{0} (0)=1$ is nonzero, while the even sidebands are weak. Therefore, the rf-field alone hardly drives the transitions between collective states with a quadratic Stark effect, so a dc field should also be present.

\begin{figure}
\includegraphics[scale=0.52]{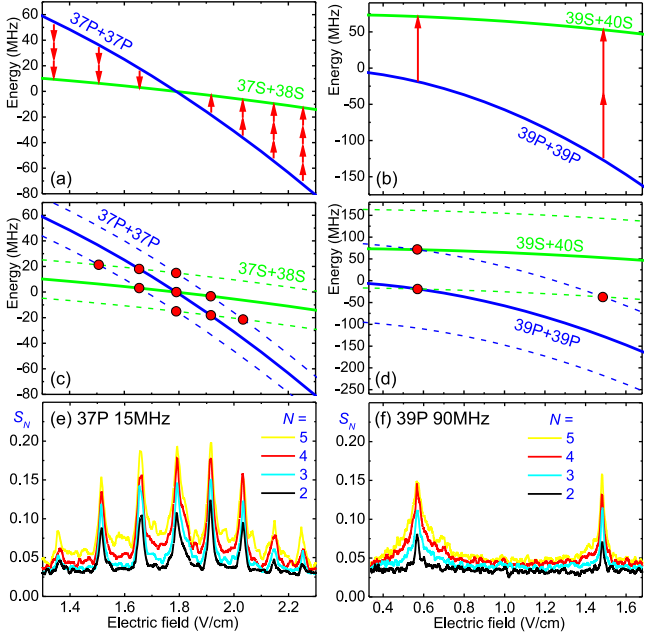}
\caption{\label{Fig6} (color online) (a) Energy levels of the initial 37\textit{P}+37\textit{P} and final 37\textit{S}+38\textit{S} collective states of two interacting Rb Rydberg atoms in an electric field. The arrows indicate rf-induced F\"orster resonances ${\rm Rb}(37P_{3/2} )+{\rm Rb}(37P_{3/2} )\to {\rm Rb}(37S)+{\rm Rb}(38S)$ of various orders at rf-frequency 15~MHz. (b) The same for the rf-induced F\"orster resonances ${\rm Rb}(39P_{3/2} )+{\rm Rb}(39P_{3/2} )\to {\rm Rb}(39S)+{\rm Rb}(40S)$ at rf frequency 90~MHz.  (c) Energy levels of the 37\textit{P}+37\textit{P} and  37\textit{S}+38\textit{S} collective states of two Rydberg atoms in an electric field in the presence of the first Floquet sidebands at $\pm$15~MHz. The red (gray) circles indicate the intersections of the Floquet sidebands corresponding to rf-assisted F\"orster resonances. (d) The same for the 39\textit{P}+39\textit{P} and 39\textit{S}+40\textit{S} collective states at rf frequency 90~MHz. (e)~Experimental record of the rf-assisted F\"orster resonances on the 37\textit{P} state for $N=2-5$ detected Rydberg atoms at a 15~MHz rf frequency and 200~mV amplitude. (f) The same for the "inaccessible" F\"orster  resonances on the 39\textit{P} state at 90~MHz and 100~mV.}
\end{figure}

Figure~6(c) shows the energy levels of the initial 37\textit{P}+37\textit{P} and final 37\textit{S}+38\textit{S} collective states of two Rydberg atoms in electric field in the presence of the first-order Floquet sidebands at $\pm$15~MHz. Their intersections correspond to rf-assisted F\"orster resonances. These resonances are presented in Fig.~6(e) on the experimental record at a 200~mV rf-amplitude. This amplitude is high enough to induce the rf transitions of up to the fourth order. Multiple rf-assisted resonances are observed due to numerous intersections of the Floquet levels in Fig.~6(c). It is remarkable that the first- and second-order resonances have nearly the same amplitude as the central resonance. This indicates that the strength of the dipole-dipole interaction at the rf-assisted F\"orster resonances is comparable with that for ordinary F\"orster resonances without the rf field.

Figures 6(d) and 6(f) show the same for the F\"orster resonance on the 39\textit{P}$_{3/2}$ state at 90~MHz and 100~mV. A narrow second-order resonance at 1.49~V/cm is well seen along with a much stronger first-order resonance at 0.57~V/cm. The first-order resonance saturates and broadens with cusped line shape as \textit{N} increases, while the second-order resonance is unsaturated and remains narrow for all \textit{N}.

Third, the amplitudes and line shapes of the rf-assisted F\"orster resonances depend on many parameters (interaction energy, interaction time, orientation of dipole moments, number of atoms, dc and rf electric-field strengths) and should be calculated numerically using the density-matrix equations to compare them with experiment. In the adiabatic approximation the F\"orster resonance detuning in Eqs.~\eqref{Eq3} adiabatically follows the Stark shift in the oscillating composite electric field \textit{F} given by Eq.~\eqref{Eq17}:

\begin{equation} \label{Eq21} 
\begin{array}{c} {\Delta (t)=\Delta _{0} +(\alpha _{nP} -\frac{1}{2} \alpha _{nS} -\frac{1}{2} \alpha _{[n+1]S} )\times } \\ \\ {\left[F_{dc} +F_{rf} \; \cos (\omega t)\right]^{2} } \end{array}.    
\end{equation} 

Given that $\Delta (t)$ is now a function of \textit{t}, Eqs.~\eqref{Eq3} cannot be solved analytically, and we applied only numerical simulations and Monte Carlo position averaging to find the theoretical spectra of two-atom rf-induced F\"orster resonances \textit{S}$_2$ with Eqs.~\eqref{Eq3}.

\subsection{Experimental results for rf-assisted F\"orster resonances}

The left-hand column in Fig.~7 presents the experimental two-atom spectra $S_{2} $ of the F\"orster resonance ${\rm Rb}(37P_{3/2} )+{\rm Rb}(37P_{3/2} )\to {\rm Rb}(37S_{1/2} )+{\rm Rb}(38S_{1/2} )$ in a~15 MHz rf field of various amplitudes at $t=2.975$~$\mu$s. Without the rf field, a single peak at 1.79~V/cm is a "true"  Stark-tuned F\"orster resonance. Application of the rf field induces additional F\"orster resonances which are rf-assisted resonances of various orders. The frequency interval between the peaks corresponds to exactly 15~MHz, taking into account the calculated polarizabilities of these Rydberg states [10]. Increasing the amplitude of the rf field leads to the appearance of the higher-order resonances. At the maximum amplitude of 300~mV, even the fifth-order resonance can be seen.

\begin{figure}
\includegraphics[scale=0.5]{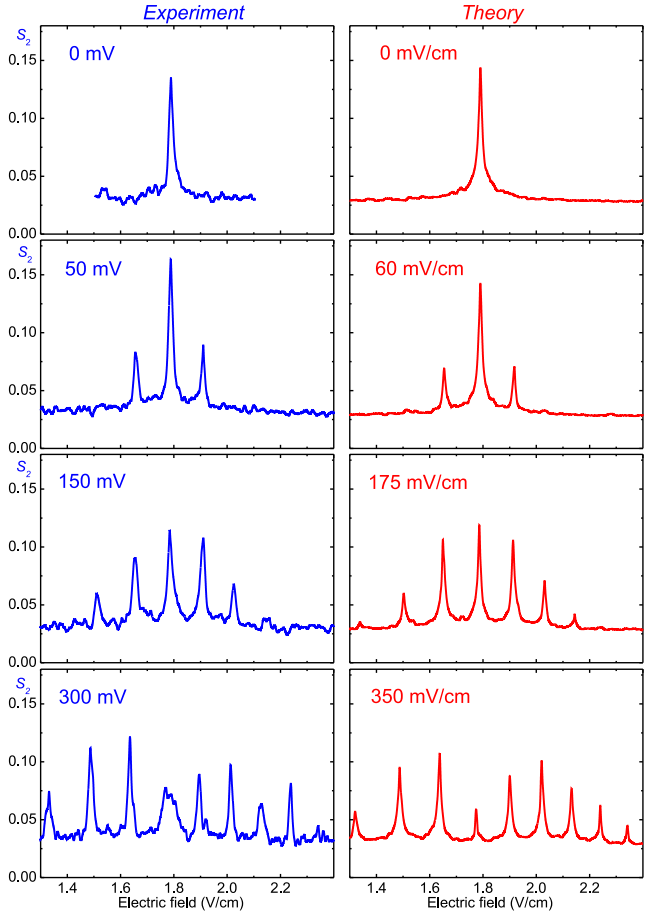}
\caption{\label{Fig7} (color online) Comparison between experimental (left-hand column) and theoretical (right-hand column) line shapes of the rf-induced F\"orster resonances Rb(37\textit{P})+Rb(37\textit{P})$\rightarrow$Rb(37\textit{S})+Rb(38\textit{S}) for two Rydberg atoms at various amplitudes of the 15 MHz rf field and interaction time $t=2.975$~$\mu$s.}
\end{figure}

The right-hand column in Fig.~7 presents the spectra, which have been numerically calculated using Eqs.~\eqref{Eq3} and \eqref{Eq21} and then spatially averaged. The fitting parameters in the theory were the volume size 30$\times$30$\times$30~$\mu$m$^3$ and the additional broadening $\Gamma /(2\pi )=0.5$~MHz. A BBR-induced background signal $S_{2} \approx 0.027$ is added to the theoretical spectra. The theoretical spectra match the experimental ones very well, both in the number and amplitude of the observed sidebands and in the line shapes at various rf amplitudes. 

However, in order to obtain this agreement, in the simulations we had to take the rf-field strength $\sim$17\% larger than the one we measured in the experiments (the experimental rf amplitudes indicated in Fig.~7 should be divided by the 1 cm spacing between our electric-field plates). We also did not manage to calculate correctly the sideband amplitudes in Fig.~7 with Eqs.~ \eqref{Eq19} and \eqref{Eq20}. This disagreement remains unclear and should be a subject of another dedicated study. One of the possible reasons can be the violation of the adiabatic approximation used in Eq.~\eqref{Eq21}.

Another discrepancy in Fig.~7 is the central peak at 1.79~V/cm and 300-mV rf amplitude. While theory predicts that it should be narrow and small, in the experiment it appears rather broad. The reason is that in the experiment, this peak is formed due to two processes. As can be seen from the electric pulse shape in Fig.~3(d), the rf pulse is applied with $\sim$100 ns delay with respect to the leading edge of the electric pulse. During that delay, there is no rf field, so a normal Fourier-broadened F\"orster resonance similar to that in Fig.~5(b) should appear. Then the rf pulse is on and this resonance is formed in the rf field. At the large amplitude of 300~mV, this field causes the ac Stark shift of the Rydberg level in the rf field, as noted below Eq.~\eqref{Eq18}. As a result, the central peak is formed by a broadened F\"orster resonance in a short dc field pulse and by a narrower shifted resonance in the rf field. 

Now we will turn to the "inaccessible" F\"orster resonance ${\rm Rb}(39P_{3/2} )+{\rm Rb}(39P_{3/2} )\to {\rm Rb}(39S_{1/2} )+{\rm Rb}(40S_{1/2} )$ which cannot be tuned by the dc electric field. Its collective energy levels in the dc electric field are shown in Fig.~6(b). The dc field alone increases the energy detuning $\Delta $ and makes the interaction between Rb(39\textit{P}) atoms less efficient. However, as we have shown in Ref.~[10], the rf field can induce transitions between collective states, so the population transfer at the F\"orster resonance occurs for the Floquet sidebands irrespective of the possibility to tune it by the dc field. In the present experiment, we applied rf field with 95~MHz frequency and 100~mV amplitude, which created the Floquet sidebands as in Fig.~6(d). 

In Figs.~8(a)-8(d), we compare the experimental first-order rf-induced F\"orster resonances recorded for two atoms at various \textit{t} to the theoretical line shapes numerically calculated with Eqs.~\eqref{Eq3} and \eqref{Eq21} and averaged over random positions of two atoms in a cubic interaction volume. A BBR-induced background signal $S_{2} \approx 0.03$ is added to the theoretical spectra. The fitting parameters in the theory were the volume size 16$\times$16$\times$16~$\mu$m$^3$, the additional broadening $\Gamma /(2\pi )=1$~MHz, and the rf-field strength 150~mV/cm (the latter is 50\% larger than the experimentally measured one; this discrepancy remains unclear). These allowed us to fit the time dependences of the experimental spectra both for the amplitude [Fig.~8(e)] and width [Fig.~8(f)] of the two-atom F\"orster resonance. The amplitude and width are measured with respect to the BBR-induced background-signal level.

The narrowest resonance appears at \textit{t}=2.975~$\mu$s. It has a cusped shape, in agreement with Eq.~\eqref{Eq15}, and its width in the electric-field scale is 18~mV/cm, corresponding to 1.1~MHz in the detuning scale, so this resonance is nearly two times narrower than that for the $37P_{3/2} $ state. The detuning scale is obtained using the calculated polarizabilities of the collective Rydberg states in Fig.~1(d). The main contribution to $\Gamma$ thus comes from the parasitic ac electric fields. That is why in the theory we have to take $\Gamma /(2\pi )=1$~MHz (i.e., two times larger than that for the $37P_{3/2} $ state), and a smaller interaction volume to fit the amplitude of the two-atom resonances in Fig.~8. 

RF-assisted inaccessible F\"orster resonances provide a way to tune the van der Waals to resonant dipole-dipole interactions and increase the interaction strength at long distances. This can be particularly useful for enhancing the dipole blockade effect [42,43] in mesoscopic Rydberg ensembles and for improving the fidelity of two-qubit quantum gates based on temporary Rydberg excitations [1,2] or of single-photon transistors based on the F\"orster resonances in Rydberg atoms [44]. For Rb \textit{nP}$_{3/2}$ states with \textit{n}=40-100, the required rf-frequencies lie in the 100$-$325~MHz range, and for \textit{nS} states with \textit{n}=70$-$120, they are in the 140$-$700~MHz range. These are reasonably low rf-frequencies, which can also be found in other alkali-metal atoms [45].

\begin{figure}
\includegraphics[scale=0.5]{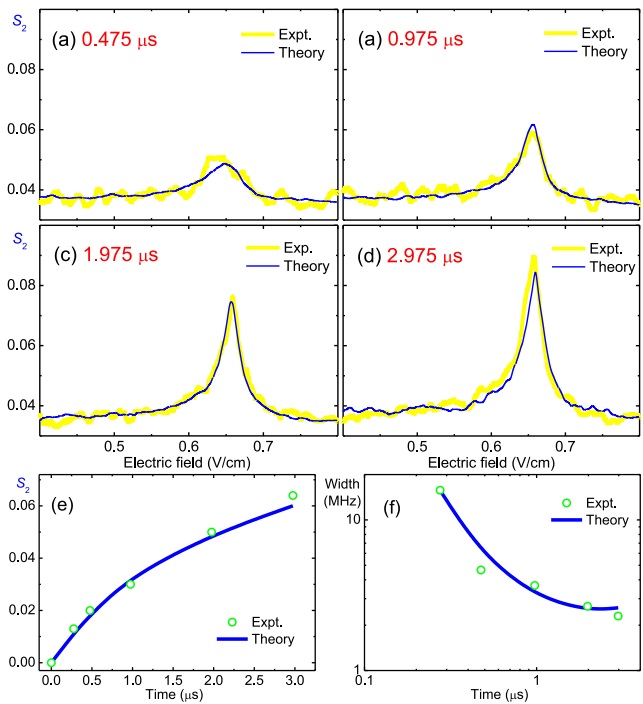}
\caption{\label{Fig8} (color online) (a)-(d) Comparison between experimental and theoretical line shapes of the inaccessible F\"orster resonance Rb(39\textit{P})+Rb(39\textit{P})$\rightarrow$Rb(39\textit{S})+Rb(40\textit{S}) induced by the 95 MHz and 100 mV rf field for two Rydberg atoms at various interaction times. (e)~Comparison between experiment and theory for the resonance amplitude at various interaction times. (f)~Comparison between experiment and theory for the resonance width at various interaction times.}
\end{figure}

\section{Conclusions}

In this paper, we have performed a detailed experimental and theoretical analysis of the line shape and time dynamics of the Stark-tuned F\"orster resonances ${\rm Rb}(nP_{3/2} )+{\rm Rb}(nP_{3/2} )\to {\rm Rb}(nS_{1/2} )+{\rm Rb}([n+1]S_{1/2} )$ for two Rb Rydberg atoms interacting in a time-varying electric field. 

In the theoretical analysis, we used an effective two-level density-matrix model that allowed us to account for the additional broadenings of the F\"orster resonances due to the unresolved hyperfine structure and parasitic ac electric fields. The analytical formulas have been obtained for the amplitude and line shape of the F\"orster resonances that correctly describe the time dynamics and dephasing of the Rabi-like population oscillations. Averaging over the random spatial positions of two Rydberg atoms in the interaction volume leads to the formation of a cusped line shape if the interaction time exceeds 1~$\mu$s and the dipole-dipole interaction is strong enough to broaden the resonance. For the strong resonance, the line shape and width are shown to depend on the interaction time. At short interaction time, the resonance width is limited by the Fourier transform of the interaction pulse.

These theoretical considerations agree well with our experimental results for two Rb Rydberg atoms interacting in a single small laser-excitation volume. We have observed both the cusp-shaped F\"orster resonances at the long interaction time and Fourier broadened resonances at the short interaction time. The time dependences of their amplitude and width are well reproduced by the numerical simulations.

Good agreement between experiment and theory confirms that any unresolved (hyperfine or Zeeman) structure or parasitic ac Stark broadening of the F\"orster resonances can be accounted for theoretically by a single parameter $\Gamma$. This significantly reduces the number of equations and simplifies the calculations of the line shapes and time dynamics of the F\"orster resonances. The time dynamics itself is represented either by the smooth exponential saturation in disordered atom ensembles, as observed in our present experiment, or by the damped Rabi-like oscillations for two frozen Rydberg atoms, as in Refs.~[31,32].

In our experiments, we have also found that the transients at the edges of the controlling electric pulses strongly affect the line shapes of the F\"orster resonances, since the population transfer at the resonances occurs on a time scale of $\sim$100~ns, which is comparable with the duration of the transients. A short-term ringing at certain frequency causes additional radio-frequency-assisted F\"orster resonances, while non-sharp edges lead to an asymmetry. 

An intentional application of the radio-frequency field induces transitions between collective states of two Rydberg atoms whose line shape depends on the interaction strengths and time. These resonances are the rf transitions between collective Floquet Rydberg states formed by the rf field due to the Stark effect. In the presence of the dc electric field, they can be induced both for F\"orster resonances which can be tuned by the dc field alone and for those which cannot be tuned. The van der Waals interaction of arbitrary Rydberg states can thus be efficiently tuned to resonant dipole-dipole interaction using the rf field.

\begin{acknowledgments}
This work was supported by the RFBR Grants No.~14-02-00680 and No.~16-02-00383, the Russian Science Foundation Grant No.~ 16-12-00028 (for laser excitation of Rydberg states), the Siberian Branch of the Russian Academy of Sciences, the EU FP7-PEOPLE-2009-IRSES Project COLIMA, and the Novosibirsk State University.
\end{acknowledgments}

\section*{APPENDIX A. Analytical solution to the density-matrix equations}

The density-matrix equations \eqref{Eq3} can be solved analytically for arbitrary $\Omega$, $\Delta$, and $\Gamma$, if these are independent of time. In order to do that, we first make the replacements

\begin{equation*}
\begin{array}{l} {\rho _{aa} +\rho _{bb} =1,} \\ {\rho _{aa} -\rho _{bb} =N,} \\ {\rho _{ab} -\rho _{ba} =Q,} \\ {\rho _{ab} +\rho _{ba} =P.} \end{array}
\end{equation*} 

\noindent Then we substitute these into Eqs.~\eqref{Eq3} and obtain

\begin{equation*}
\begin{array}{l} {\dot{N}=2\sqrt{2} \; i\Omega Q,} \\ {\dot{P}=-i\Delta Q-\Gamma P/2,} \\ {\dot{Q}=-i\Delta P-\Gamma Q/2+2\sqrt{2} \; i\Omega N.} \end{array} 
\end{equation*} 

\noindent After several substitutions we come to a single differential equation

\begin{equation*}
\dddot{N}+\Gamma \ddot{N}+(8\Omega ^{2} +\Delta ^{2} +\Gamma ^{2} /4)\dot{N}+4\Gamma \Omega ^{2} N=0. 
\end{equation*} 

\noindent Then we seek the solution as $N\sim {\rm e}^{\mu t} $and obtain the cubic equation,

\begin{equation*}
\mu ^{3} +\Gamma \mu ^{2} +(8\Omega ^{2} +\Delta ^{2} +\Gamma ^{2} /4)\mu +4\Gamma \Omega ^{2} =0. 
\end{equation*} 

\noindent This equation has three roots, which are found analytically with the following sequence of equations taken from the mathematical handbooks:

\begin{equation*}
\begin{array}{l} 
{S=8\Omega ^{2} +\Delta ^{2} +\Gamma ^{2} /4,} \\ 
{T=4\Gamma \Omega ^{2},} \\ 
{A=S-\Gamma ^{2} /3=8\Omega ^{2} +\Delta ^{2} -\Gamma ^{2} /12,} \\ 
{B=2\Gamma ^{3} /27-\Gamma S/3+T=\Gamma (4\Omega ^{2} -\Delta ^{2} -\Gamma ^{2} /36)/3, }\\
{D=(A/3)^{3} +(B/2)^{2},} \\ 
{U=(-B/2+\sqrt{D} )^{1/3},} \\ 
{V=-(B/2+\sqrt{D} )^{1/3}. }
\end{array} 
\end{equation*} 

\noindent Finally, the three roots are

\begin{equation*}
\begin{array}{l} 
{\mu _{1} =U+V-\Gamma /3,} \\
{\mu _{2} =-(U+V)/2+i\sqrt{3} (U-V)/2, }\\ 
{\mu _{3} =-(U+V)/2-i\sqrt{3} (U-V)/2. }
\end{array} 
\end{equation*} 

Then we seek \textit{N} in the form

\begin{equation*}
N=c_1\; {\rm e}^{\mu _{1} t} +c_2\; {\rm e}^{\mu _{2} t} +c_3\; {\rm e}^{\mu _{3} t}  
\end{equation*} 

\noindent with the initial conditions $N(0)=1;\; \dot{N}(0)=0;\; \ddot{N}(0)=-8\Omega ^{2} $. These conditions give us the final exact analytical solution for the coefficients,

\begin{equation*}
\begin{array}{l} 
{c_1=\displaystyle \frac{\mu _{2} \mu _{3} -8\Omega ^{2} }{(\mu _{1} -\mu _{2} )(\mu _{1} -\mu _{3} )}\;, }\\ \\ 
{c_2=\displaystyle \frac{\mu _{1} \mu _{3} -8\Omega ^{2} }{(\mu _{1} -\mu _{2} )(\mu _{3} -\mu _{2} )}\;,} \\ \\ 
{c_3=\displaystyle \frac{\mu _{1} \mu _{2} -8\Omega ^{2} }{(\mu _{1} -\mu _{3} )(\mu _{2} -\mu _{3} )}\;. } 
\end{array}
\end{equation*} 

These formulas allow us to find \textit{N} analytically and thus provide the exact analytical solution to Eqs.~\eqref{Eq3} for arbitrary $\Omega$, $\Delta$, $\Gamma$, and \textit{t}. The two-atom signal is then calculated as

\begin{equation*}
S_{2} =(1-N)/4 \;.
\end{equation*} 

The exact formulas are rather complex and cannot be presented in a clearly understandable way. Therefore, in Secs. IV~B and IV~C we consider the analytical solutions only for some particular cases where the formulas are reduced to short expressions.

\section*{APPENDIX B. Spatial averaging of the F\"orster resonance amplitude}

For two interacting Rydberg atoms randomly positioned in a single laser-excitation volume, the spatial averaging of the F\"orster resonance amplitude given by Eqs.~\eqref{Eq6} or \eqref{Eq7} can be performed analytically in the same way as in Ref.~[22]. The averaging is done using the nearest-neighbor probability distribution [33,34] with the average distance between nearest-neighbor atoms $R_{0} \approx \left[3/(4\pi n_{0} )\right]^{1/3} $ at volume density \textit{n}$_0$,

\begin{equation*}
P(R)={\rm e}^{-R^{3} /R_{0}^{3} } 3R^{2} /R_{0}^{3} \; . 
\end{equation*}

The amplitude is averaged as 

\begin{equation*}
<S_{2} (\Delta =0)>=\int _{0}^{\infty }S_{2} (\Delta =0)P(R)dR \;. 
\end{equation*}

\noindent This integral is finite because Eqs.~\eqref{Eq6} and \eqref{Eq7} are finite even at $R\to 0$ when $\Omega \to \infty $. Its approximate analytical solution can be found if we use a conventional removal of the orientation-dependent part in Eq.~\eqref{Eq2}; that is, we consider a "scalar" dipole-dipole interaction. We remove the $3Z_{ab}^{2} /R_{ab}^{5} $ term in Eq.~\eqref{Eq2}, so the interaction strength is now given by $1/R_{ab}^{3} $ or $1/R^{3} $. Its validity for disordered atom ensembles is grounded by the fact that mutual atom orientations are effectively averaged in such ensembles. Then we introduce 

\begin{equation*}
\Omega _{0} =\frac{\sqrt{2} d_{1} d_{2} }{4\pi \varepsilon _{0} \hbar R_{0}^{3} }  
\end{equation*}

\noindent as the orientation-averaged reduced interaction matrix element at the average distance $R_{0} $. Here, $d_{1} $ and $d_{2} $ are the dipole moments of transitions ${nP_{3/2} \left({\rm M}_{J} =1/2\right) } \to {nS_{1/2} \left({\rm M}_{J} =1/2\right) } $ and ${nP_{3/2} \left({\rm M}_{J} =1/2\right) } \to {(n+1)S_{1/2} \left({\rm M}_{J} =1/2\right) } $. We also define a dimensionless integration variable $x=(R/R_{0} )^{3} $. 

At the weak dipole-dipole interaction, the integral to be found for the averaged amplitude of the F\"orster resonance is 

\begin{equation*}
\left\langle S_{2}^{weak} (\Delta =0)\right\rangle \approx \frac{1}{4} \left[1-\int _{0}^{\infty }{\rm exp}\left(-{\tfrac{8\Omega _{0}^{2} t}{x^{2} \; \Gamma }} -x\right)dx \right]. 
\end{equation*}

\noindent The integrand in this expression is zero at $x\to 0$ and at $x\to \infty $, but it has a maximum at $x_{0} =\left(16\Omega _{0}^{2} t/\Gamma \right)^{1/3} $. This maximum is $\exp \left[-1.5\left(16\Omega _{0}^{2} t/\Gamma \right)^{1/3} \right]$, while the width of the integrand in the \textit{x} scale is of the order of 1. We therefore assumed that with a certain accuracy, the integral can be estimated as $\alpha \exp \left[-\left(\beta \Omega _{0}^{2} t/\Gamma \right)^{1/3} \right]$. Here $\alpha ,\; \beta \sim 1$ are the fitting parameters, which can be found by numerical simulations. In fact, we should take $\alpha =1$ to make the amplitude zero at \textit{t}=0. Then our numerical simulation for a cubic interaction volume gave us $\beta \approx 0.44$, so the average amplitude for the weak interaction is approximately

\begin{equation*}
<S_{2}^{weak} (\Delta =0)>\approx \frac{1}{4} \left(1-{\rm e}^{-\left[0.44\Omega _{0}^{2} \; t/\Gamma \right]^{1/3} } \right). 
\end{equation*}

At the strong dipole-dipole interaction, the integral to be found for the averaged amplitude of the F\"orster resonance is 

\begin{equation*}
\left\langle S_{2}^{strong} (\Delta =0)\right\rangle \approx \frac{1}{4} \left[1-{\rm e}^{-\Gamma t/4} \int _{0}^{\infty }\cos \left({\tfrac{2\; \Omega _{0} t}{x}} \right)\; {\rm e}^{-x} dx \right]. 
\end{equation*}

\noindent In this case, we first do a trigonometric substitution $\cos \left(2\Omega _{0}^{2} t/x\right)=1-2\sin ^{2} \left(\Omega _{0}^{2} t/x\right)$ and obtain a modified formula

\begin{equation*}
\begin{array}{l}
{\left\langle S_{2}^{strong} (\Delta =0)\right\rangle \approx \frac{1}{4} \left[1-{\rm e}^{-\Gamma t/4} +\right. } \\ \\{\left. {\rm 2e}^{-\Gamma t/4} \int _{0}^{\infty }\sin ^{2} \left({\tfrac{\Omega _{0} t}{x}} \right)\; {\rm e}^{-x} dx \right]}. 
\end{array} 
\end{equation*}

An approximate analytical solution may be found as in Ref.~[22] if we note that the main contribution to the integral comes from the very short distances, where the rapidly oscillating squared sine function can be replaced with its average value 1/2. Then the integration should start at $x=0$ and stop at some point $x_{0} \approx \alpha \Omega _{0} t$, where $\alpha \sim 1$, because the integrand exponentially drops at $x>x_{0} $. Again, $\alpha $ is a fitting parameter, which can be found by numerical simulations. Our simulations for a cubic interaction volume gave us $\alpha \approx 0.55$, so the average amplitude for the strong interaction is approximately

\begin{equation*}
<S_{2}^{strong} (\Delta =0)>\approx \frac{1}{4} \left(1-{\rm e}^{-0.55\Omega _{0} \; t-\Gamma t/4} \right). 
\end{equation*}

\section*{APPENDIX C. Spatial averaging of the F\"orster resonance resonance line shape}

The F\"orster resonance line shape can be averaged analytically in the same way as the amplitude. The main requirement is that the integral over \textit{R} must not diverge at $R\to 0$ when $\Omega \to \infty $. Therefore, the averaging can be done with Eqs.~\eqref{Eq13} or \eqref{Eq14} that correctly describe the saturation of F\"orster resonances. But, in fact, Eq.~\eqref{Eq14} cannot be averaged analytically at $\Delta \ne 0$ due to its complicated dependence on \textit{R}. Only its Lorentz pre-factor can be averaged as in Ref.~[23], but it does not provide the dependence on the interaction time. 

Therefore, in our averaging, we use Eq.~\eqref{Eq13} with $\Gamma \ne 0$. It is distinguished by the fact that upon spatial averaging, the third term in Eq.~\eqref{Eq13} averages to nearly zero at $\Delta \ne 0$ and $\Gamma t>1$, while the second term remains significant at $\Gamma \ne 0$ and fully defines the averaged line shape. The averaged line shape to be found is 

\begin{equation*}
\left\langle S_{2}^{strong} \right\rangle \approx \frac{1}{4} \left[1-\int _{0}^{\infty }\frac{\Delta ^{2} x^{2} }{4\Omega _{0}^{2} +\Delta ^{2} x^{2} } {\rm e}^{-\frac{2\Omega _{0}^{2} \Gamma t}{4\Omega _{0}^{2} +\Delta ^{2} x^{2} } -x} dx \right] .
\end{equation*}

\noindent At large $\Delta $ we can neglect $4\Omega _{0}^{2} $ in the denominators, and the integral reduces to

\begin{equation*}
\left\langle S_{2}^{strong} \right\rangle \approx \frac{1}{4} \left[1-\int _{0}^{\infty }{\rm e}^{-\frac{2\Omega _{0}^{2} \Gamma t}{\Delta ^{2} x^{2} } -x} dx \right] .
\end{equation*}

\noindent The integrand in this expression is zero at $x\to 0$ and at $x\to \infty $, but it has a maximum at $x_{0} =\left(4\Omega _{0}^{2} \Gamma t/\Delta ^{2} \right)^{1/3} $. This maximum is $\exp \left[-1.5\left(4\Omega _{0}^{2} \Gamma t/\Delta ^{2} \right)^{1/3} \right]$, and the width of the integrand in the \textit{x} scale is of the order of 1. We therefore suggested that, to some precision, the integral can be estimated as $\alpha \exp \left[-\left(\beta \Omega _{0}^{2} \Gamma t/ \Delta^2\right)^{1/3} \right]$, with $\alpha ,\; \beta \sim 1$ being the fitting parameters, which can be found by numerical simulations. In fact, we should take $\alpha =1$ to make the amplitude zero at \textit{t}=0. Then the average line shape for the weak interaction is approximately

\begin{equation*}
\left\langle S_{2}^{strong} \right\rangle \approx \frac{1}{4} \left[1-\exp \left(-\left\{\frac{\beta \Omega _{0}^{2} \Gamma t}{\Delta ^{2} } \right\}^{1/3} \right)\right] .
\end{equation*}

Our numerical simulations have shown that this formula works well at large $\Delta $, but it does not work correctly near $\Delta =0$. We assumed that it can work better if we modify it phenomenologically as follows:

\begin{equation*}
\left\langle S_{2}^{strong} \right\rangle \approx \frac{1}{4} \left[1-\exp \left(-\left\{\frac{0.44\; \Omega _{0}^{2} \Gamma t}{a^{2} \Delta ^{2} +\Gamma ^{2} } \right\}^{1/3} \right)\right] .
\end{equation*}

The fitting coefficients 0.44 and \textit{a} have been obtained by comparing Eq.~\eqref{Eq15} to the numerical simulations of Eqs.~\eqref{Eq3} averaged over a cubic interaction volume for the F\"orster resonance Rb(37\textit{P})+Rb(37\textit{P})$\rightarrow$Rb(37\textit{S})+Rb(38\textit{S}). Depending on $\Omega_0$, $\Gamma$, and \textit{t}, the values of \textit{a} vary in the range from 2 to 4, as discussed in Sec. IV C.


\begin{thebibliography}{10}

\bibitem{1} M.~Saffman, T.~G.~Walker, and K.~M$\o$lmer, Rev. Mod. Phys. \textbf{82}, 2313 (2010).

\bibitem{2} I.~I.~Ryabtsev, I.~I.~Beterov, D.~B.~Tretyakov, V.~M.~Entin, and E.~A.~Yakshina, Physics $-$ Uspekhi \textbf{59}, 196 (2016).

\bibitem{3} K.~A.~Safinya, J.~F.~Delpech, F.~Gounand, W.~Sandner, and T.~F.~Gallagher, Phys. Rev. Lett. \textbf{47}, 405 (1981).

\bibitem{4} P.~Pillet, D.~Comparat, M.~Muldrich, T.~Vogt, N.~Zahzam, V.~M.~Akulin, T.~F.~Gallagher, W.~Li, P.~Tanner, M.~W.~Noel, and I.~Mourachko, in \textit{Decoherence, Entanglement and Information Protection in Complex Quantum Systems}, edited by V.~M.~Akulin \textit{et al.} (Springer, New York, 2005), p.411.

\bibitem{5} A.~Tauschinsky, C.~S.~E.~van~Ditzhuijzen, L.~D.~Noordam, and H.~B.~van~Linden van~den~Heuvell, Phys. Rev. A \textbf{78}, 063409 (2008).

\bibitem{6} D.~B.~Tretyakov, I.~I.~Beterov, V.~M.~Entin, I.~I.~Ryabtsev, and P.~L.~Chapovsky, J. Exper. Theor. Phys. \textbf{108}, 374 (2009).

\bibitem{7} I.~I.~Ryabtsev, D.~B.~Tretyakov, I.~I.~Beterov, and V.~M.~Entin, Phys. Rev. Lett. \textbf{104}, 073003 (2010).

\bibitem{8} D.~B.~Tretyakov, I.~I.~Beterov, V.~M.~Entin, I.~I.~Ryabtsev, N.~N.~Bezuglov, and E.~Arimondo, J. Exper. Theor. Phys. \textbf{114}, 14 (2012).

\bibitem{9} V.~M.~Entin, E.~A.~Yakshina, D.~B.~Tretyakov, I.~I.~Beterov, and I.~I.~Ryabtsev, J. Exper. Theor. Phys. \textbf{116}, 721 (2013).

\bibitem{10} D.B.Tretyakov, V.M.Entin, E.A.Yakshina, I.I.Beterov, C.Andreeva, and I.I.Ryabtsev, Phys. Rev. A \textbf{90}, 041403(R) (2014).

\bibitem{11} I.~I.~Ryabtsev and I.~M.~Beterov, Phys. Rev. A \textbf{61}, 063414 (2000).

\bibitem{12} I.~I.~Ryabtsev, D.~B.~Tretyakov, and I.~I.~Beterov, J. Phys. B \textbf{36}, 297 (2003). 

\bibitem{13} S.~Westermann, T.~Amthor, A.~L.~de~Oliveira, J.~Deiglmayr, M.~Reetz-Lamour, and M.~Weidem\"uller, Eur. Phys. J. D \textbf{40}, 37 (2006).

\bibitem{14} J.~Nipper, J.~B.~Balewski, A.~T.~Krupp, B.~Butscher, R.~L\"ow, and T.~Pfau, Phys. Rev. Lett. \textbf{108}, 113001 (2012).

\bibitem{15} P.~J.~Tanner, J.~Han, E.~S.~Shuman, and T.~F.~Gallagher, Phys. Rev. Lett. \textbf{100}, 043002 (2008).

\bibitem{16} K.~C.~Younge, A.~Reinhard, T.~Pohl, P.~R.~Berman, and G.~Raithel, Phys. Rev. A \textbf{79}, 043420 (2009).

\bibitem{17} J.~R.~Veale, W.~Anderson, M.~Gatzke, M.~Renn, and T.~F.~Gallagher, Phys. Rev. A \textbf{54}, 1430 (1996).

\bibitem{18} I.~I.~Ryabtsev, D.~B.~Tretyakov, I.~I.~Beterov, and V.~M.~Entin, Phys. Rev. A \textbf{76}, 012722 (2007); Erratum: Phys. Rev. A \textbf{76}, 049902(E) (2007).

\bibitem{19} W.~R.~Anderson, J.~R.~Veale, and T.~F.~Gallagher, Phys. Rev. Lett. \textbf{80}, 249 (1998).

\bibitem{20} I.~Mourachko, D.~Comparat, F.~de~Tomasi, A.~Fioretti, P.~Nosbaum, V.~M.~Akulin, and P.~Pillet, Phys. Rev. Lett. \textbf{80}, 253 (1998).

\bibitem{21} B.~Sun and F.~Robicheaux, Phys. Rev. A \textbf{78}, 040701(R) (2008).

\bibitem{22} I.~I.~Ryabtsev, D.~B.~Tretyakov, I.~I.~Beterov, V.~M.~Entin, and E.~A.~Yakshina,  Phys. Rev. A \textbf{82}, 053409 (2010).

\bibitem{23} B.~G.~Richards and R.~R.~Jones, Phys. Rev. A \textbf{93}, 042505 (2016).

\bibitem{24} J.~M.~Kondo, D.~Booth, L.~F.~Goncalves, J.~P.~Shaffer, and L.~G.~Marcassa, Phys. Rev. A \textbf{93}, 012703 (2016).

\bibitem{25} B.~Pelle, R.~Faoro, J.~Billy, E.~Arimondo, P.~Pillet, and P.~Cheinet, Phys. Rev. A \textbf{93}, 023417 (2016).

\bibitem{26} H.~Park, T.~F.~Gallagher, and P.~Pillet, Phys. Rev. A \textbf{93}, 052501 (2016).

\bibitem{27} S.~Stenholm, \textit{Foundations of Laser Spectroscopy} (Dover, Mineoloa, NY, 2005).

\bibitem{28} I.~I.~Beterov, I.~I.~Ryabtsev, D.~B.~Tretyakov, and V.~M.~Entin, Phys. Rev. A \textbf{79}, 052504 (2009).

\bibitem{29} G.~S.~Agarwal, Phys. Rev. Lett. \textbf{37}, 1383 (1976).

\bibitem{30} K.~Wodkiewicz, Phys. Rev. A \textbf{19}, 1686 (1979).

\bibitem{31} S.~Ravets, H.~Labuhn, D.~Barredo, L.~Beguin, T.~Lahaye, and A.~Browaeys, Nature Physics\textit{ }\textbf{10},  914 (2014).

\bibitem{32} S.~Ravets, H.~Labuhn, D.~Barredo, T.~Lahaye, and A.~Browaeys, Phys. Rev. A \textbf{92}, 020701(R) (2015).

\bibitem{33} P.~Hertz, Math. Ann. \textbf{67}, 387 (1909). 

\bibitem{34} S.~Chandrasekhar, Rev. Mod. Phys. \textbf{15}, 1 (1943).

\bibitem{35} P.~Pillet, R.~Kachru, N.~H.~Tran, W.~W.~Smith, and T.~F.~Gallagher, Phys. Rev. A \textbf{36}, 1132 (1987).

\bibitem{36} C.~S.~E.~van~Ditzhuijzen, A.~Tauschinsky, and H.~B.~van~Linden van~den~Heuvell, Phys. Rev. A \textbf{80}, 063407 (2009).

\bibitem{37} Y.~Yu, H.~Park, and T.~F.~Gallagher, Phys. Rev. Lett. \textbf{111}, 173001 (2013).

\bibitem{38} M.~G.~Bason, M.~Tanasittikosol, A.~Sargsyan, A.~K.~Mohapatra, D.~Sarkisyan, R.~M.~Potvliege, and C.~S.~Adams, New J. Phys. \textbf{12}, 065015 (2010).

\bibitem{39} S.~A.~Miller, D.~A.~Anderson, and G.~Raithel, New J. Phys. \textbf{18}, 053017 (2016).

\bibitem{40} S.~Yoshida, C.~O.~Reinhold, J.~Burgdorfer, S.~Ye, and F.~B.~Dunning, Phys. Rev. A \textbf{86}, 043415 (2012).

\bibitem{41} V.~Zhelyazkova and S.~D.~Hogan, Phys. Rev. A \textbf{92}, 011402(R) (2015).

\bibitem{42} M.~D.~Lukin, M.~Fleischhauer, R.~Cote, L.~M.~Duan, D.~Jaksch, J.~I.~Cirac, and P.~Zoller, Phys. Rev. Lett. \textbf{87}, 037901 (2001).

\bibitem{43} D.~Comparat and P.~Pillet, J. Opt. Soc. Am. B \textbf{27}, A208 (2010).

\bibitem{44} D.~Tiarks, S.~Baur, K.~Schneider, S.~D\"urr, and G.~Rempe, Phys. Rev. Lett. \textbf{113}, 053602 (2014).

\bibitem{45}   J.~H.~Gurian, P.~Cheinet, P.~Huillery, A.~Fioretti, J.~Zhao, P.~L.~Gould, D.~Comparat, and P.~Pillet, Phys. Rev. Lett. \textbf{108}, 023005 (2012).

\end{thebibliography}
\end{document}